\documentclass[a4paper]{article}

\usepackage{INTERSPEECH2021}
\usepackage[linesnumbered,ruled,vlined]{algorithm2e}
\usepackage{multirow}
\usepackage[
    hidelinks,
]{hyperref}

\makeatletter
\def\Hline{
  \noalign{\ifnum0=`}\fi\hrule \@height 1pt \futurelet
  \reserved@a\@xhline}
\makeatother
\newcommand{\Tau}{\scalebox{1.2}{$\tau$}}

\usepackage{float}

\usepackage{caption} 
\newcommand{\xcdot}[2]{[#1_1, #1_2, \cdots, #1_{#2}]^{\intercal}}

\newcommand{\nofont}[1]{{\fontfamily{qcr}\selectfont #1}}
\newcommand{\what}[1]{\widehat #1}

\newcommand{\loss}{\mathcal{L}}
\newcommand{\real}{\mathbb{R}}

\newcommand{\bX}{\boldsymbol{X}}
\newcommand{\bx}{\boldsymbol{x}}
\newcommand{\by}{\boldsymbol{y}}
\newcommand{\bC}{\boldsymbol{C}}
\newcommand{\bZ}{\boldsymbol{Z}}
\newcommand{\bH}{\boldsymbol{H}}

\newcommand{\blank}{\text{\nofont{blank}}}
\newcommand{\ctc}{\text{ctc}}
\newcommand{\ar}{\text{ar}}
\newcommand{\mlm}{\text{mlm}}
\newcommand{\att}{\text{att}}
\newcommand{\jca}{\text{jca}}
\newcommand{\stud}{\text{student}}

\newcommand{\tmask}{\text{mask}}
\newcommand{\tobs}{\text{obs}}
\newcommand{\mask}{\textless\nofont{MASK}\textgreater~}

\newcommand{\fkd}{\loss_{\text{F-KD}}}
\newcommand{\skd}{\loss_{\text{S-KD}}}
\newcommand{\kd}{\text{KD}}
\newcommand{\encoder}{\text{Encoder}}
\newcommand{\decoder}{\text{Decoder}}
\newcommand{\genc}{\gamma_{\text{enc}}}
\newcommand{\gdec}{\gamma_{\text{dec}}}

\title{Knowledge Transfer and Distillation from Autoregressive to Non-Autoregressive Speech Recognition}

\name{Xun Gong, Zhikai Zhou, Yanmin Qian$^\dagger$
\thanks{
    $^\dagger$corresponding author
}
}
\address{
MoE Key Lab of Artificial Intelligence, AI Institute \\
X-LANCE Lab, Department of Computer Science and Engineering \\
Shanghai Jiao Tong University, Shanghai, China
}
\email{\{gongxun,zhikai.zhou,yanminqian\}@sjtu.edu.cn}

\begin{document}

\maketitle
\begin{abstract}
Modern non-autoregressive~(NAR) speech recognition systems aim to accelerate the inference speed; however, they suffer from performance degradation compared with autoregressive~(AR) models as well as the huge model size issue.
We propose a novel knowledge transfer and distillation architecture that leverages knowledge from AR models to improve the NAR performance while reducing the model's size.
Frame- and sequence-level objectives are well-designed for transfer learning.
To further boost the performance of NAR, a beam search method on Mask-CTC is developed to enlarge the search space during the inference stage.
Experiments show that the proposed NAR beam search relatively reduces CER by over 5\% on AISHELL-1 benchmark with a tolerable real-time-factor~(RTF) increment.
By knowledge transfer, the NAR student who has the same size as the AR teacher obtains relative CER reductions of 8/16\% on AISHELL-1 dev/test sets, and over 25\% relative WER reductions on LibriSpeech test-clean/other sets.
Moreover, the $\sim$9x smaller NAR models achieve $\sim$25\% relative CER/WER reductions on both AISHELL-1 and LibriSpeech benchmarks with the proposed knowledge transfer and distillation.
\end{abstract}

\section{Introduction}

In recent years, the performance of automatic speech recognition~(ASR) has been greatly improved by sequence-to-sequence modeling, such as connectionist temporal classification~(CTC)~\cite{graves2006ctc}, recurrent neural network transducer~(RNN-T)~\cite{battenberg2017rnnt}, and attention-based encoder-decoder~(AED)~\cite{hori2017joint}.
Many of the earlier researches have focused on autoregressive (AR) modeling, which generates the token sequence using a left-to-right probabilistic chain rule.
Despite their great performance, such AR models require $L$-step incremental model calculations to generate $L$ tokens, resulting in high inference latency and considerable computational costs.

From another aspect, non-autoregressive (NAR) modeling generates the token sequence in constant steps, and removes the chain-rule assumption.
CTC~\cite{graves2006ctc} plays an essential role in recent NAR researches.
Modern NAR approaches outperform CTC by leveraging alignments~(align-based) and the output token sequence~(token-based).
Based on the joint CTC/attention architecture~\cite{hori2017joint}, Mask-CTC~\cite{higuchi2020mask} utilizes the (conditional) masked language model~((C)MLM) decoder to refine the CTC token sequence.
\cite{higuchi2021improved} proposes two auxiliary tasks to solve the length prediction problem that occurred in Mask-CTC.
From another point of view, CTC alignment shows its advantage for building NAR models in Align-Refine~\cite{chi2021alignrefine}, CASS-NAT~\cite{fan2021improved} and AL-NAT~\cite{wang2022alignmentlearning}.
Moreover, the self-supervised model wav2vec2.0~\cite{baevski_2020_wav2vec_20_framework} has achieved promising results by CTC.

However, there are two major challenges remaining for NAR modeling:
First, NAR models converge slowly and perform poorly compared to state-of-the-art~(SOTA) AR models~\cite{higuchi2020mask,deng2022improving}.
Second, although NAR models are often favored in resource-constrained situations for fast inference speed and high accuracy~\cite{higuchi2020mask}, the large model scale and computational cost limit the application of NAR modeling.
Knowledge distillation~(Transfer learning) is typically used to solve such a problem by teaching a smaller student model~\cite{gou2021knowledge}.
To be specific, the student aims to mimic the soft targets provided by the well-trained teacher using the Kullback-Leibler divergence~(KLD)~\cite{huang2018knowledgea,takashima2019investigation,munim2019sequencelevel}.
Nevertheless, when applying knowledge distillation on non-autoregressive ASR, the poor NAR teacher limits the improvement.

In this paper, we propose a novel architecture by transferring and distilling the knowledge from an autoregressive~(AR) teacher to a non-autoregressive~(NAR) student together with a beam-search decoding method to boost the performance of non-autoregressive modeling.
Firstly, we introduce a beam-search decoding method to enlarge the search space for the (conditional) masked language model~((C)MLM) decoder~\cite{ghaz2019mask}.
Then, we extend the knowledge distillation technique by transferring the AR teacher's knowledge to NAR in two distillation levels, therefore improving NAR students' performance.
The encoder distillation is conducted following our previous setup~\cite{huang2018knowledgea}.
For the decoder distillation, we develop the frame- and sequence-level distillation from the attention-based autoregressive model into Mask-CTC.
The distillation loss is customized for token-based NAR models, so that the NAR decoder can benefit from the AR decoder.

The structure of the paper is organized as follows:
In Section~\ref{sec:related}, the attention-based autoregressive model and non-autoregressive Mask-CTC are briefly introduced.
In Section~\ref{sec:method}, we present the proposed beam search method for Mask-CTC and the knowledge distillation from AR to NAR ASR.
In Section~\ref{sec:expr}, experimental results and analysis are given on the Mandarin AISHELL-1 and English LibriSpeech datasets.
Finally, the conclusion is drawn in Section~\ref{sec:conc}.

\section{Autoregressive and Non-autoregressive ASR}
\label{sec:related}

Basically, end-to-end ASR models map speech features $\bX = \xcdot{\bx}{T}, \bx_t \in \real^F$ to a token sequence $\by = \xcdot{y}{L}, y_l \in \mathcal{U}$, where $F$ is the feature dimension and $\mathcal{U}$ denotes the vocabulary set.

Traditional attention-based autoregressive~(AR) ASR models~\cite{vaswani2017attention,gulati2020conformer} firstly encode speech features $\bX$ into a hidden representation $\bH$: $\bH = \text{Encoder} (\bX)$, and then compose it with previous tokens $\by_{<l}$ to estimate the posterior $p(y_l | \by_{<l}, X)$:
\begin{align}
    p_{\ar}(y_l | \by_{<l}, \bH) &= \text{Decoder}(\by_{<l}, \bH) \label{eq:attention},
\end{align}
and the whole sequence probability is
\begin{equation}
     p_{\ar}(\by | \bH) = p_{\ar}(y_1 | \bH) \prod_{l=2}^{L} p_{\ar}(y_l | \by_{<l}, \bH).
\end{equation}
During inference, the AR model generates hypothesis $\hat \by$ token-by-token.

Connectionist temporal classification~(CTC)~\cite{graves2006ctc} is one of the earliest non-autoregressive~(NAR) method, which introduces a many-to-one function $\eta$ from the frame-level alignment $\bZ = \xcdot{z}{T}, z_t \in \mathcal{U} \cup \{\blank\}$ to the token sequence $\by$, by merging same labels and removing the $\blank$ in $\bZ$. The sequence probability is represented as:
\begin{align}
\label{eq:ctc}
    p_{\ctc}(\by | \bH) = \sum_{\bZ \in \eta^{-1}(\by)} \prod_t p_{\ctc}(z_t | \bH),
\end{align}
where $\eta$ is a many-to-one function from $\bZ$ to $\by$.
During inference, greedy CTC predicts the alignment by selecting the tokens with the highest probability for each step.

Mask-CTC~\cite{higuchi2020mask}, which is a popular instantiation of NAR ASR, is actually a refinement of CTC results via the conditional masked language model~(MLM)~\cite{ghaz2019mask}.
During training, the groundtruth $\by$ are randomly replaced by the special token \mask, and the MLM decoder predicts masked tokens $\by_{\tmask}$, conditioned on the observed tokens $\by_{\tobs} = \by \setminus \by_{\tmask}$:
\begin{align}
\label{eq:mlm}
    p_{\mlm} (\by_{\tmask} | \by_{\tobs}, \bH) = \prod_{y \in \by_{\tmask}} p_{\mlm} (y | \by_{\tobs}, \bH).
\end{align}
During inference, the output is initialized by CTC greedy decoding, and low-confidence tokens are substituted by \mask based on pre-defined threshold $p_{\text{thr}}$. 
After that, masks are filled using easy-first algorithm:
fill all masks in $\lceil N/K \rceil$ iterations, where $N$ denotes the total number of \mask and each iteration predicts top-$k$ tokens with the highest confidence guided by MLM:
\begin{align}
    \hat{\bC} &= \mathop{\arg\max}_{\substack{\bC \subset \by_{\tmask}, \\ |\bC| = k}} \prod_{c \in \bC} p_{\mlm} (c | \by_{\tobs}, \bH), \label{eq:infer_mlm1}
\end{align}
where $k = \begin{cases} N \mod K, &\text{the last iteration} \\ K, &\text{otherwise} \end{cases}$, $\bC$ is the candidate set of \mask tokens and $\ \what{\by'_{\tobs}} = \by_{\tobs} + \hat{\bC}\ $ is the updated result after mask filling.

The joint CTC/attention architecture~\cite{hori2017joint,higuchi2020mask} is widely used in modern AR and NAR ASR models, with a multi-task learning based loss function:
\begin{align}
\label{eq:joint_loss}
    \loss_{\jca} = \alpha \loss_{\ctc} + (1-\alpha) \loss_{\att},
\end{align}
where $\alpha \in [0,1]$ is a hyperparameter, $\loss_{\att} = \loss_{\ar}$ for AR ASR, and $\loss_{\att} = \loss_{\mlm}$ for NAR ASR, respectively.

\section{Proposed Methods}
\label{sec:method}

In this section, we introduce: (1) the proposed beam search method for NAR, (2) the distillation architecture transferring knowledge from AR to NAR ASR.

\subsection{Beam Search for NAR ASR}
\label{sec:beam_search}

Inspired by joint and rescoring decoding~\cite{hori2017joint}, we design a beam-search decoding method to enlarge the search space for the MLM decoder.
The procedure is shown in Algorithm~\ref{alg:beam}.
$\Omega_1$ is a sorted queue to be updated at the beginning of one iteration, $\Omega_0$ stores the final $\Omega_1$ after one iteration.
During each iteration, a $B$-size beam is preserved, and the number of updated tokens is fixed and computed by $k$~(i.e. $|\what{y_{\tobs}^{i+1}}| = |\what{y_{\tobs}^{i}}| + k$).
Top-$B$ candidates are selected, according to the log domain posterior probability and Equation~\ref{eq:infer_mlm1}.

\begin{algorithm}
\caption{Beam Search on NAR MLM Decoding}
\label{alg:beam}

$\hat{\by} \leftarrow$ greedy CTC \;
$\hat{\by}_{\tmask}$ and $\hat{\by}_{\tobs}^0$ by $p_{thr}$ \;
$\Omega_0 \leftarrow \{ \hat{\by}_{\tobs}^0\}$: a list to store accepted hypotheses \;
\For{each iteration $i = 1,2,\cdots,\lceil N / K \rceil$}{
    $\Omega_1 \leftarrow \varnothing$: $\Omega_1$ is a sorted queue with $B$ length \;
    
    $k = \begin{cases} N \mod K, &\text{the last iteration,} \\ K, &\text{otherwise.} \end{cases}$ \;
    
    \For{$\by_{\tobs} \in \Omega_0$}{
        Get top-$B$ candidate sets $\what{\by_{\tobs}^{'1}}, \cdots, \what{\by_{\tobs}^{'B}}$ by the posterior in Equation~\ref{eq:infer_mlm1} \;
        Push $\what{\by_{\tobs}^{'1}}, \cdots, \what{\by_{\tobs}^{'B}}$ into $\Omega_1$ \;
    }
    
    $\Omega_0 = \Omega_1$ \;
}

\Return $\mathop{\arg\max}\limits_{\hat{\by} \in \Omega_0} \hat{\by}$ \;
\end{algorithm}

\subsection{Knowledge Transfer and Distillation from Autoregressive to Non-Autoregressive ASR}

\begin{figure}[ht]
    \centering
    \includegraphics[width=\linewidth]{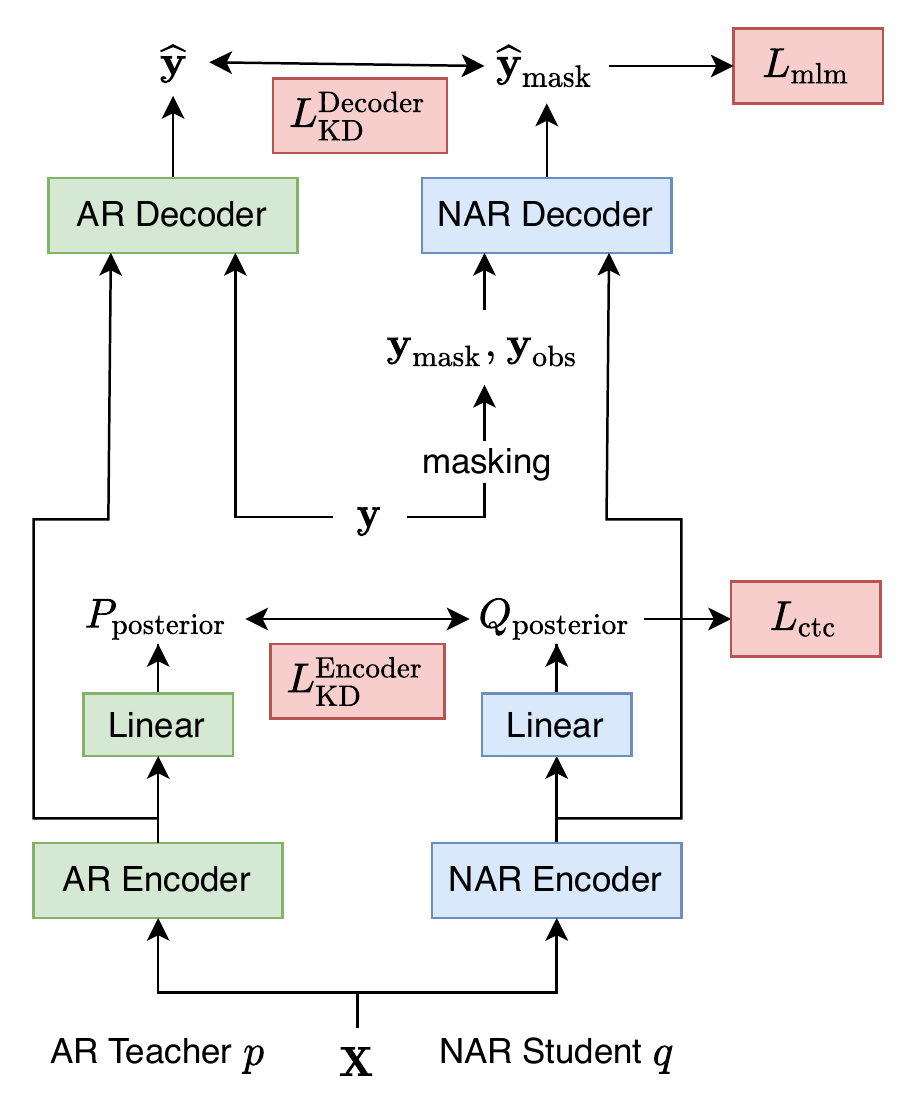}
    \caption{Overview of proposed knowledge distillation from autoregressive to non-autoregressive ASR.}
    \label{fig:model}
\end{figure}

As previously stated, knowledge distillation performance on NAR is constrained owing to the poor performance of the NAR teacher.
Here, we propose knowledge transfer and distillation from autoregressive~(AR) to non-autoregressive~(NAR) ASR, pushing the limits of NAR.

Firstly, we introduce two types of knowledge distillation techniques based on Kullback-Leibler divergence~(KLD): $KLD(P, Q) = \sum_i P_i \log (P_i / Q_i)$, where $P, Q$ are the teacher and student output distributions, respectively.

\textbf{Frame-level knowledge distillation} as a basic distillation criterion is formulated as below:
\begin{align}
    \label{eq:frame_kd}
    \fkd &= - \sum_{t=1}^T \sum_{c \in \mathcal{U}} P_t(c) \log Q_t(c),
\end{align}
where $P_t(c)$ and $Q_t(c)$ are the posterior probabilities of token $c$ at timestamp $t$ of teacher $P$ and student $Q$. $\bH$, $\by_{\tobs}$, and $\by_{<t}$ are the conditions of the above probabilities, but omitted for simplicity.
$P_t(c) \log P_t(c)$ is omitted in computing KLD loss due to the frozen teacher during training.

\textbf{Sequence-level knowledge distillation} is another distillation criterion:
\begin{align}
    \label{eq:seq_kd}
    \skd = - \sum_{\what{\by^i} \in \Tau} P(\what{\by^i}) \log Q(\what{\by^i})
\end{align}
where $\what{\by^i}$ is the hypotheses from teacher model, $\Tau$ is the set of all possible sequences and similar omission as in Equation~\ref{eq:frame_kd}.
Using such sequence-level knowledge distillation is unaffordable, as we are approximating an exponentially-sized sequence distribution $\Tau$.
Similar to MWER~\cite{prabhavalkar2017minimum} training, N-best candidates set $\Omega$ is accessed by beam search, and then $P(\what{\by}^i)$ can be approximated by:
\begin{align}
    \label{eq:seq_kd_approx}
    P(\what{\by^i}) \simeq \frac{P(\what{\by^i})} {\sum_{\what{\by^j} \in \Omega} P(\what{\by^j})}.
\end{align}

And then we can achieve the knowledge distillation loss by:
\begin{align}
    \label{eq:loss_kd}
    \loss_{\kd} = \beta_F \fkd + \beta_S \skd,
\end{align}
where $\beta_F, \beta_S$ are hyper-parameters for frame-level knowledge distillation loss $\fkd$ and sequence-level $\skd$, respectively.

As shown in Figure~\ref{fig:model}, the proposed knowledge distillation methods are divided into two parts: the first part is the distillation after the encoder, and the second part is the distillation after the decoder.
The encoder distillation $\loss_{\kd}^{\encoder}$ is done after the linear layer of the encoder, which has $\fkd$ and $\skd$ similar to literature \cite{huang2018knowledgea}.
The decoder distillation is setup as follows.
For frame-level distillation, only $\by_{\tmask}$ positions are selected, so the objective function is normalized by the number of \mask tokens:
\begin{align}
    \label{eq:loss_lkd_dec}
    \fkd^{\decoder} = - \sum_{t \in \by_{\tmask}} \sum_{c\in \mathcal{U}} \frac{P_t(c | \by_{<t}, \bH)}{|\by_{\tmask}|} \log Q_t(c | \by_{\tobs}, \bH).
\end{align}

For sequence-level distillation, approximate probability $P'$ from N-best $\Omega$ is used:
\begin{align}
    \label{eq:loss_skd_dec}
    \skd^{\decoder} = - \sum_{\hat{\by} \in \Omega} \frac{P'(\hat{\by})}{|\by_{\tmask}|} \log Q( \by_{\tmask} |  \by_{\tobs}, \bH)).
\end{align}

And the final loss is then:
\begin{align}
    \label{eq:final_loss}
    \loss = \loss_{\jca}^{\stud} + \genc \loss_{\kd}^{\encoder}+ \gdec \loss_{\kd}^{\decoder},
\end{align}
where $\genc, \gdec$ are weight coefficients for encoder and decoder knowledge distillation.

\section{Experiments}
\label{sec:expr}
\begin{table*}[ht]
    \caption{Knowledge transfer and distillation performance on AISHELL1 corpus~(CER)~(\%) and Librispeech test corpus~(WER)~(\%), `I+D' is the sum of insertion and deletion errors, and `A' is the total CER/WER. `\#Param' has brackets with `XS', `S', `M', `L', which are mentioned in Table~\ref{tab:model_size}. `Same size' means NAR has the same model scale as AR, `Smaller' means NAR is $\sim$9x smaller than AR.}
    \label{tab:expr_aishell}
    \label{tab:expr_libri}
    \centering
    \begin{tabular}{@{\extracolsep{4pt}} l | l | c cccc | c cccc @{}}
    \Hline
        \multirow{3}{*}{Method} & \multirow{3}{*}{Criterion} & \multicolumn{5}{c|}{AISHELL1} & \multicolumn{5}{c}{Librispeech Test} \\
        \cline{3-7}\cline{8-12}
         &&\multirow{2}{*}{\#Param} & \multicolumn{2}{c}{Dev} & \multicolumn{2}{c|}{Test~} & \multirow{2}{*}{\#Param} & \multicolumn{2}{c}{clean~} & \multicolumn{2}{c}{other} \\
         & && I+D & A & I+D & A && I+D & A & I+D & A \\
    \hline \hline
        W2V2-CTC~\cite{deng2022improving,baevski_2020_wav2vec_20_framework} & - & 95M & -  & 4.8 & - & 5.3 & 95M & - & 2.7 & - & 8.0 \\
        CASS-NAT v2~\cite{fan2021improved} & - & - & - & 4.9 & - & 5.4 & - & - & 3.1 & - & 7.2 \\
        AL-NAT~\cite{wang2022alignmentlearning} & - & 71.3M & - & 4.9 & - & 5.3 & $\sim$115M & - & 3.2 & - & 7.4 \\
    \hline
        AR & $\loss_{jca}$ & 46.8M~(M) & 0.2 & 4.5 & 0.2 & 4.9 & 115.0M~(L) & 0.2 & 2.8 & 1.3 & 6.6 \\
    \hline
        NAR~(Same size) & $\loss_{jca}$ & 46.8M~(M) & 0.4 & 5.4 & 0.5 & 6.4 &115.0M~(L) & 0.6 & 4.1 & 2.0 & 10.2 \\
        & +$\fkd$ & - & 0.2 & \textbf{5.1} & 0.3 & 5.6 &- & 0.5 & 3.5 & 1.6 & 8.4 \\
        & ++$\skd$ & - & 0.2 & \textbf{5.0} & 0.2 & \textbf{5.4} &- & 0.4 & \textbf{3.3} & 1.4 & \textbf{7.8} \\
    \hline
        NAR~(Smaller) & $\loss_{jca}$ & 6.5M~(XS) & 0.4 & 7.2 & 2.1 & 8.5 & 12.4M~(S) & 0.8 & 5.1 & 2.4 & 12.4 \\
                 & +$\fkd$ & - & 0.2 & \textbf{6.1} & 0.5 & 6.9 &- & 0.4 & 3.8 & 1.8 & 10.3  \\
                 & ++$\skd$& - & 0.2 & \textbf{6.0} & 0.3 & \textbf{6.5} &- & 0.3 & \textbf{3.7} & 1.4 & \textbf{9.2}  \\
    \Hline
    \end{tabular}
\end{table*}

\subsection{Datasets}

Our experiments are conducted on the Mandarin AISHELL-1~\cite{bu2017AISHELL1} and the English LibriSpeech corpora~\cite{panayotov2015LibriSpeech}. 
AISHELL-1 contains a 150h training set, with a development~(dev) and test set for evaluation, while LibriSpeech has a 960h training set, with test-clean/other~(test~c/o) used for tests.
We report the character error rate~(CER) on AISHELL-1, word error rate~(WER) on LibriSpeech, respectively.

\subsection{Model description}

For acoustic feature extraction, 80-dimensional mel filterbank~(Fbank) features are extracted with global level cepstral mean and variance normalization~(CMVN).
When it comes to data augmentation, speed perturbation is applied only for AISHELL-1 and SpecAugment~\cite{park2019specaugment} for both datasets, respectively.
For text modeling, 5000 English Byte Pair encoding~(BPE)~\cite{kudo2018sentencepiece} subword units are adopted for English, and 4233 characters for Mandarin.
The baseline follows the recipe of ESPnet v2~\cite{watanabe2018espnet}, a 12-layer conformer encoder with four times down-sampling and a 6-layer transformer decoder.
The weight $\alpha$ for CTC module is fixed to $0.3$.

For knowledge transfer and distillation, we firstly train a new NAR student model from scratch with $\fkd$ for 80 epochs.
The hyper-parameters are set to $\beta_F = 1.0, \beta_S = 0, \genc = 0.5,\gdec = 0.3$.
Then we fine-tune the distillation procedure by adding $\skd$, and the tuning parameters are set as $\beta_F = 1.0, \beta_S = 1.0, \genc = 0.5,\gdec = 0.5$ with 20 epochs.
In Equation~\ref{eq:seq_kd},~\ref{eq:seq_kd_approx}, we use beam-size $|\Omega| = 10$, which is consistent with the decoding hyper-parameters in AR model.

\begin{table}[ht]
    \caption{Model hyper-parameters for different AR and NAR Conformer scales at L, M, X and XS.}
    \label{tab:model_size}
    \centering
    \begin{tabular}{c|c c c c}
    \Hline
        Model & L & M & S & XS \\
    \hline
    \hline
        \#Params~(M)    &115.0&46.8 &12.4 & 6.5 \\
        Attention Dim   
        & 512 & 256 & 128 & 128\\
        Attention Heads & 8   & 4   & 4   & 4  \\
        Inner-dim       &2048 &2048 &1024 & 256\\
    \Hline
    \end{tabular}
\end{table}

Different NAR student model sizes are explored in Table~\ref{tab:model_size}, identified as large~(L), medium~(M), small~(S), and extremely small~(XS).
The AR teacher model keeps L size for LibriSpeech, and M-size for AISHELL-1.

In the inference stage, no language model is used during the following experiments.
Model parameters are averaged over the last 5 checkpoints.
For autoregressive models, joint CTC/attention one-pass decoding~\cite{hori2017joint} is used with beam size 10, and the score interpolation of CTC is 0.3.
For non-autoregressive Mask-CTC decoding, we follow the beam decoding method in Section~\ref{sec:beam_search}, with beam $B = 10$, the threshold $p_{\text{thr}} = 0.99$ and $K = 2$, for AISHELL-1 and LibriSpeech corpus.

\subsection{Results with the NAR Beam Decoding}

As proposed in Section~\ref{sec:beam_search}, we firstly evaluate the beam search performance with real time factor~(RTF) in Table~\ref{tab:expr_beam}.
RTF is computed using Intel-Xeon E5-2690 CPU with a single core at test set.
To be consistent with literature~\cite{higuchi2020mask}, the NAR~(M) model speeds up AR~(M)~\cite{hori2017joint} model by more than 10x, as the RTF is 0.58 for AR~(M) and 0.31 for AR~(S).
Without too much degradation of inference speed~(1.5x as slow as `Beam1'), the beam decoding method achieves better performance compared with the greedy~(Beam1) one by 5\%$\sim$9\% relative WER reduction on the test set.
As the beam size $B$ grows, the rate of improvement decreases.

\begin{table}[ht]
    \caption{Non-autoregressive Mask-CTC Performance (CER) on AISHELL-1 corpus. Real time factor~(RTF) are reported for the test set.}
    \label{tab:expr_beam}
    \centering
    \begin{tabular}{l ccc}
    \Hline
        Decoding & Dev~(\%) & Test~(\%) & RTF \\
    \hline

    \multicolumn{4}{l}{\textit{Non-Autoregressive~(M)}} \\
        \quad Beam1 & 5.4 & 6.7 & 0.051 \\ %
        \quad Beam5 & 5.4 & 6.4 & 0.064 \\
        \quad Beam10 & 5.4 & 6.4 & 0.072 \\
        
    \multicolumn{4}{l}{\textit{Non-Autoregressive~(XS)}} \\
        \quad Beam1 & 7.6 & 9.2 & 0.017 \\
        \quad Beam5 & 7.3 & 8.6 & 0.023 \\
        \quad Beam10 & 7.2 & 8.5 & 0.027 \\
    \Hline
    \end{tabular}
\end{table}

\subsection{Results on Knowledge Transfer and Distillation}

Table~\ref{tab:expr_aishell} compares the knowledge transfer distillation and other modern AR and NAR models on AISHELL-1 and LibriSpeech datasets to validate the performance.

\textbf{AISHELL-1: } As shown in Table~\ref{tab:expr_aishell}, the teacher AR model obtains more than 24\% relative reduction on CER compared with NAR~(M), and 40\% with NAR~(XS).
After knowledge distillation, the NAR~(M) with `+$\fkd$' achieves 8\% and 16\% relative CER reduction on dev and test sets respectively, and the one based on `+$\fkd$' with `++$\skd$' shows a further CER reduction on test set by 15\%.
The student results achieve competitive performance~(5.0\%/5.4\% CER) to the state-of-the-art NAR models like CASS-NAT~\cite{fan2021improved} or AL-NAT~\cite{wang2022alignmentlearning}.
Similar results are obtained for distilled NAR~(XS) as 18\%/25\% CER reduction on two evaluation sets.

\textbf{LibriSpeech: } Table~\ref{tab:expr_libri} shows the performance comparison on the large LibriSpeech corpus.
The AR~(L) is adopted as teacher model while NAR~(L,S) as student model.
The observations are consistent with that in Table~\ref{tab:expr_aishell}, $\fkd$ and $\skd$ further boost the performance of NAR Mask-CTC model at L~(3.3/7.8\% WER) and S~(3.7/9.2\% WER) scale by $\sim$25\% relative WER reduction.
However, due to the limits of the AR teacher, the insertion and deletion error rate is high on AR~(L).

Results show that such our method narrows the gap between AR and NAR, with the improvement being significantly greater in the more difficult evaluation set (i.e. test set in AISHELL-1, test-other in LibriSpeech).
After knowledge transfer and distillation, the length error problem is greatly alleviated compared with the original NAR model owing to the high prediction accuracy of AR teacher.
Moreover, both $\fkd$ and $\skd$ attribute to reducing insertion and deletion errors~(`I+D'), pushing the length error problem~\cite{higuchi2020mask} to its limits at 0.2\% CER for `I+D' in AISHELL-1 and 1.4\% for LibriSpeech test-other.
Meanwhile, the NAR student model performs comparable results with other state-of-the-art NAR methods, including wav2vec2-CTC~\cite{baevski_2020_wav2vec_20_framework,deng2022improving}, Improved CASS-NAT~\cite{fan2021improved} and AL-NAT~\cite{wang2022alignmentlearning}.

\section{Conclusions}
\label{sec:conc}

In this paper we propose a novel knowledge transfer and distillation architecture that leverages knowledge from AR models to improve NAR performance while reducing the model size.
To further boost the performance of NAR, we propose a beam search method on Mask-CTC, which enlarges the search space during inference stage.
Experiments demonstrate that NAR beam search obtains relative 5\% reduction in AISHELL-1 dataset with a tolerable RTF increment.
For knowledge distillation, most results achieve over 15\% relative CER/WER reduction on large and smaller NAR modeling.
Future works are going to explore the generalization of knowledge distillation from AR to NAR.
Different non-autoregressive like CASS-NAT~\cite{fan2021improved} and AL-NAT~\cite{wang2022alignmentlearning} might be explored with external non-autoregressive language models.
Hidden feature distillation~\cite{gou2021knowledge} is also a valuable extension of this paper.

\section{Acknowledgment}

This work is supported in part by China NSFC projects under Grants 62122050 and 62071288, and in part by Shanghai Municipal Science and Technology Major Project under Grant 2021SHZDZX0102.
The authors would like to thank Tian tan and Wangyou Zhang for discussion.

\bibliographystyle{IEEEtran}

\bibliography{mybib}

\end{document}